\documentclass[12pt]{article}
\usepackage{arxiv}

\title{Joint estimation of insurance loss development factors using Bayesian hidden Markov models}
\author[]{Conor Goold\thanks{conor@ledgerinvesting.com}}
\affil[]{Ledger Investing, Inc., New York, USA}

\begin{document}
\maketitle

Loss development modelling is the actuarial practice of predicting the 
total \textit{ultimate} losses incurred on a set of policies once all 
claims are reported and settled. This poses a challenging prediction 
task as losses frequently take years to fully emerge from 
reported claims, and not all claims might yet be reported.
Loss development models often estimate a set of 
\textit{link ratios} from insurance loss triangles, which are multiplicative 
factors transforming losses at one time point to ultimate. However, 
link ratios estimated using classical methods typically underestimate 
ultimate losses and cannot be extrapolated outside the domains of the 
triangle, requiring extension by \textit{tail factors} from another model. 
Although flexible, this two-step process relies on
subjective decision points that might bias inference.
Methods that jointly estimate `body’ link ratios and 
smooth tail factors offer an attractive 
alternative. This paper proposes a novel application of Bayesian 
hidden Markov models to loss development modelling, where discrete, 
latent states representing body and tail processes are automatically 
learned from the data. The hidden Markov development model 
is found to perform comparably to, and frequently better than, the 
two-step approach, as well as a latent change-point model,
on numerical examples and industry datasets.


\section{Introduction}
\label{sec:introduction}
Actuarial science, founded in the seventeenth century,
is one of the oldest areas
of applied statistics, 
and insurance continues to pose
many interesting statistical questions
about risk and uncertainty.
A core task in actuarial science is
\textit{loss development} or \textit{loss reserving}.
Loss development is the practice of
predicting the total 
payments due on all 
claims expected from
a set of insurance policies \citep{englandverrall2002,zhang2012}.
Thus, loss development requires
forecasting not only future payments for
claims already reported to insurance companies 
but payments for claims yet to be reported.
In this way, insurance is unique in that 
the eventual cost of 
an insurance policy
is unknown at the time of sale
because insurance policies cover, for the most part,
prospective accidents only.
Together with the time it takes to report claims
to insurance companies, and the time it can
take for an insurance company to arrive
at a final claim loss payment amount,
these features of insurance make knowing
the total losses an insurance company will
be liable to pay significantly uncertain. 

The total losses due for claims
from a pool of insurance policies
are typically referred to as
the \textit{ultimate} losses, and are composed
of the loss payments already made on opened
claims (paid losses), plus an amount estimated by actuaries
for payments on claims yet to be made (loss reserves).
This latter amount is frequently referred to as
\textit{incurred but not reported} (IBNR), and
includes future payments on
claims already reported (`incurred but not enough reported'),
as well as payments for claims not yet reported (`incurred but not
yet reported').
In non-life insurance, 
such as property and casualty insurance,
IBNR can be substantial as it may
take years for all claims on a set of policies to be reported and settled.
Therefore, the accurate estimation of IBNR is key to evaluating an insurance company's
performance and solvency \citep{beard1960,bornhuetter1972,friedland2010,englandverrall2002,wuthrich2008}.
Transparent assessment of the uncertainty in IBNR 
is also critical to the legal responsibilities
of insurers and reinsurers, such as for the Solvency II directive
\citep{england2019,frohlich2018,munroe2018}.

Estimation of loss reserves 
has received considerable attention from actuaries, econometricians,
and statisticians for decades 
\citep[e.g. for some key developments, see][]{bornhuetter1972,clarke1974,
taylor1977,taylor1983,mack1993,barnett2000,englandverrall2001,englandverrall2002,
taylor2003,wuthrich2008}. 
Loss development models are typically applied to sets of aggregated and tabulated
data known as \textit{loss triangles}, where the rows represent
intervals of time when accidents occurred, known as experience periods,
and the columns represent evaluation points, known as development periods or lags,
showing the cumulative or incremental cost of those claims reported in each accident year
(see Figure \ref{fig:schematic} for a worked example).
The data structure represents an upper triangular matrix, hence
the name `loss triangle', and it is the task of loss development modelling 
to complete, and potentially extend, the
lower diagonal to produce predictions of ultimate loss.
From early deterministic and 
algebraic approaches \citep[e.g.][]{scurfield1968,
bornhuetter1972,clarke1974,taylor1977}, loss development modelling has
advanced to utilise a wide, and growing, variety of statistical methods, 
including parametric and non-parametric regressions \citep{mack1994,englandverrall2001,
englandverrall2002,lally2018}, Bayesian estimation \citep{englandverrall2002,dealba2002,
zhang2012,meyers2015}, differential equations \citep{gesmann2020}, and 
neural networks and machine learning methods 
\citep{kunce2017,kuo2019,almudafer2022}.

A key inferential quantity from any loss development
model is the set of multiplicative factors transforming 
the losses at development period one to
the ultimate losses at development period $\infty$,
known as the \textit{loss development factors} or \textit{link ratios}. 
While many models include link ratios as an explicit parameter to be 
inferred, notably the family of chain ladder methods
\citep[e.g.][]{mack1993,englandverrall2002},
others derive link ratios as a generated quantity \citep{englandverrall2001,meyers2015}. 
Ideally, link ratios for cumulative losses will smoothly decline over time towards unity,
indicating that the losses are approaching their ultimate value at later development
periods.
In practice, however, link ratios often include periods of volatility, particularly
for early development periods, and may further encode systematic
and non-systematic effects across experience or 
development periods (e.g. the influence of the Covid-19 pandemic).
Depending on the degree of volatility and the amount of data available,
link ratios estimated using popular methodologies
may be insufficient to infer the ultimate losses for each experience period.
For instance, at the latest development period in the data, link ratios
may not be nearing unity, suggesting further development is yet to come.
Consequently, estimation of ultimate losses will require extending the link ratios
to outside the domains of the focal triangle to include additional
link ratios, known as \textit{tail factors}, that make assumptions
about how the losses will continue to develop in the future.

Like the loss development literature discussed above, 
tail factor estimation has had its own expansive history
of deterministic and stochastic methods \citep{tailfactors2013}, and is of particular
importance in `long-tailed' lines of business, such as workers' compensation or general
liability, where growth of loss payments towards ultimate loss might be slow or
volatile, 
even at relatively late development periods.
Of the various approaches used to calculate tail factors, many 
use a second model fit to a portion
of the focal triangle or to a subset of the link ratios
that conveys how the triangle may behave
in the tail, close to ultimate loss \citep{tailfactors2013}.
These models typically 
infer a parametric, monotonically
increasing growth curve of losses from the training data, 
such as various forms of
inverse power curves \citep[e.g.][]{sherman1984,evans2015,clark2017}.
The link ratios derived from this tail model are then appended to
the link ratios estimated from the primary loss development model
to produce predictions an
arbitrary number of development periods into the future.
For clarity, the primary link ratios will be referred to
in this paper as the `body' link ratios to distinguish them from 
link ratios estimated in the tail.

The two-step process of estimating body and tail link ratios
includes a number of 
subjective decisions that may not be optimal. 
Namely, it necessitates manually selecting a development 
period where the tail link ratios should take over from
the body link ratios. It may also involve manually selecting portions of
loss triangles to be used as training data that best match
body and tail model assumptions.
The choice of development periods to use in each case
can have important implications for predictions of ultimate
loss.
For instance, including periods
of non-monotonic growth into tail model training
data would bias
tail factors, and ultimate losses, unreasonably high. 
Similarly, predicting
ultimate losses using tail factors applied too early
in a triangle
could miss important development features at later
development periods and bias ultimate losses too low. 
Moreover, the manual selection of training data windows
and body-to-tail switch-over points
are difficult to reproduce, time consuming and
opens analysts to many `researchers degrees of freedom' \citep{simmons2011}.

Methods that simultaneously estimate body
and tail link ratios 
present an attractive alternative
to traditional two-step processes. However, only a few
solutions have been proposed.
\cite{englandverrall2001} presented a generalised
additive model for smooth estimation of loss development
factors that can be extrapolated to points futher
than the existing data. Although flexible and able to
integrate different functional forms and covariates
of loss development processes, this approach
still necessitates careful selection of training data
to include in the model so that regions of volatility
do not bias loss development curves. Generalised additive models
further require selecting the family of splines and
number of knots to apply, which may lead to another
set of decisions analysts must act on.
\cite{zhang2012}, alternatively,
implemented a hierarchical Bayesian logistic growth curve model to
cumulative loss data. However, fitting a single parametric
curve that assumes monotonicity in expectation might under-estimate
systematic volatility in portions of the triangle that
analysts do not want to label simply as residual noise.
Finally, \cite{verrall2012} use reversible jump Markov
chain Monte Carlo to combine a Bayesian chain ladder model,
applied before some body-to-tail switch-over point, with an exponential
decay process in the tail, and allow the model to infer
where the switch-over should occur. Additionally, 
\cite{verrall2015} demonstrate how the same model
can be estimated in a Bayesian model averaging context.
Although this approach is arguably the most flexible,
it requires either bespoke sampling algorithms (i.e. reversible
jump Monte Carlo) or
multiple model fits (for model averaging purposes),
which may dissuade analysts and researchers.

This paper proposes the use of hidden Markov models
to simultaneously estimate body and tail link ratios
in a single model from a variety of loss triangles. 
Hidden Markov models are 
primarily discrete mixture models postulating an unknown 
latent state
underlying and generating patterns of observed data,
and have found a multitude of applications, 
from speech recognition \citep[e.g.][]{rabiner1989}
to animal behaviour \citep[e.g.][]{leos2017}.
Indeed, Markov processes have been previously used
in micro-level claim 
modelling \citep[e.g.][]{hesselager1994} but
have not been applied to aggregate insurance
loss triangles.
As described in this paper, hidden Markov 
development models decompose 
a loss triangle into a sequence of body and tail
processes.
Hidden Markov models are easily fit
in existing, open-source software, and can cater
for complex data-generating assumptions,
such as understanding the impact of covariates
or including non-parametric patterns of loss development 
\citep{englandverrall2001},

Below, the hidden Markov model is formulated
and validated with simulated examples and 
tested on multiple data sources. 
The model is compared to both a traditional
two-step modelling process and a slightly
more parsimonious latent
change-point model.
Models are fit using Bayesian estimation
and following a modern Bayesian
workflow \citep{gelman2020}.
All code and datasets to 
reproduce the results
are accessible at the Github repository
\url{https://github.com/LedgerInvesting/paper-hidden-markov-development}.

\section{Hidden Markov development model}
Consider a typical loss development context, where
a large set of homogeneous insurance risks (e.g. private
car insurance policies) are aggregated
into a loss triangle with
cumulative loss amounts denoted
$\mathcal{Y}$, defined by:

\begin{equation}
	\mathcal{Y} = \{y_{ij} : i = 1, ..., N; j = 1, ..., N - i + 1\}
\end{equation}
where $i = 1, ..., N$ denotes experience periods 
indexing all claims occurring during period $i$, and $j = 1, ..., M$
denotes development periods. 
Most frequently, $N = M$.
For a particular point in time,
development information for period $i$ is only known up to period 
$j = N - i + 1$,
and therefore $\mathcal{Y}$ represents the left upper diagonal of
the loss triangle. The complementary, lower diagonal 
triangle, denoted $\tilde{\mathcal{Y}}$,
with $\frac{N (M - 1)}{2}$ entries 
for a square matrix,
is unknown and the goal of prediction.

\begin{figure}[t!]
    \centering
    \includegraphics[scale=0.9]{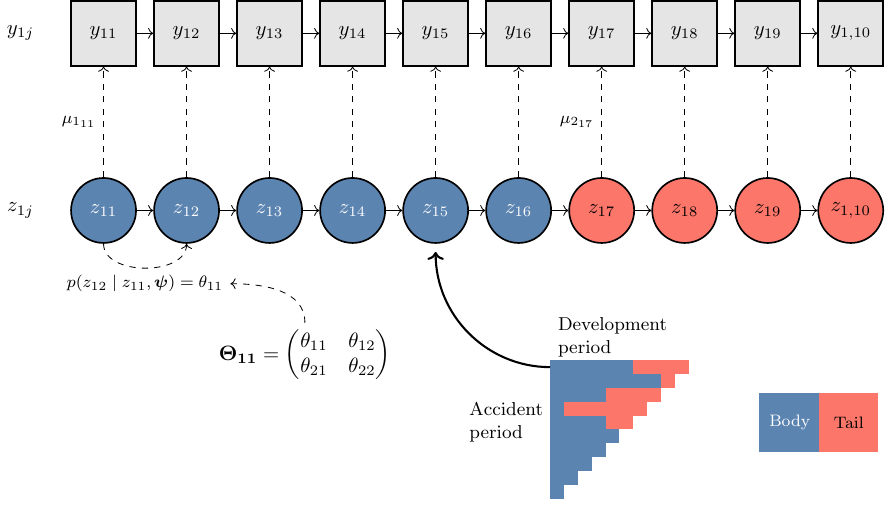}
    \caption{
        A schematic of the hidden Markov model process.
        A loss triangle of observed data is shown
        with 10 accident periods. Observed losses
        at each development period and accident period
        is generated from one of
        two processes: the body (blue)
        and or tail (orange). The dynamics of the
        body and tail states vary over accident periods.
        For the first accident period, $y_{1j}$,
        the hidden Markov process is illustrated
        across development periods $j = 1, ..., 10$.
        Observed losses are shown as grey
        squares and are assumed generated from the latent
        discrete random variable $z_{1j}$ (circles) via
        emission distributions.
        Latent states transition according
        to the probabilities
        in matrix $\Theta_{1j}$, depending on parameters
        $\bm{\psi}$, and $\mu_{1_{1j}}$ and $\mu_{2_{1j}}$
        represent the body and tail state emission distributions.
   }
	\label{fig:schematic}
\end{figure}

The generative process considered here, show in
Figure \ref{fig:schematic},
is that cumulative losses
in $\mathcal{Y}$ develop according to $i$) a period that is
characterised by largely, but not strictly, monotonically 
increasing losses 
(the body), followed by $ii$) a period of smooth
growth to ultimate (the tail). Rather than
treat estimation of body and tail as a two-step process,
the hidden Markov development model introduces a
latent, discrete state $\bm{z} = (z_{11}, z_{12}, ..., z_{ij}, ..., z_{NM})
\in \{1, ...,  K\}$, with $K = 2$, that takes value
$z_{ij} = 1$ if the body process generated the losses in the $i$th
accident period and $j$th development period, and
$z_{ij} = 2$ if the tail process generated the losses.
The latent state at one time point, $z_{ij}$, is dependent
on the state at the previous time point, $z_{ij-1}$,
and subsequent states are connected via a state transition
matrix, $\Theta_{ij}$.
Depending on the latent state and any other parameters
$\bm{\phi}$, the likelihood, $p(y_{ij} \mid z_{ij}, \bm{\phi})$,
is given by suitable \textit{emission} distributions.
Emission distributions are the observation data distributions,
which in loss development modelling are typically
positive-bound, continuous probability density
functions, such as lognormal or Gamma.
In this paper, all models use lognormal distributions
as the likelihood distribution.

The hidden Markov development model can be
written generally as:

\begin{align}
	\begin{split}
	\label{eq:hmm}
	y_{ij} &\sim \begin{cases}
        \mathrm{Lognormal}(\mu_{1_{ij}}, \sigma_{ij}) \quad z_{ij} = 1\\
        \mathrm{Lognormal}(\mu_{2_{ij}}, \sigma_{ij}) \quad z_{ij} = 2\\
	\end{cases}\\
    z_{ij} &\sim \mathrm{Categorical}(\Theta^{z_{ij-1}}_{ij})\\
    \Theta_{ij} &= \begin{pmatrix}
        \pi & 1 - \pi\\
        \nu & 1 - \nu
    \end{pmatrix}\\
    \log \frac{\pi}{1 - \pi} &\sim \mathrm{Normal}(0, 1)\\
    \log \frac{\nu}{1 - \nu} &\sim \mathrm{Normal}(0, 1)\\
	\end{split}
\end{align}
The cumulative losses are positive-bound, 
lognormally-distributed random variates,
with log-scale location $\mu_{1_{ij}}$ if the process is in the body
or $\mu_{2_{ij}}$ if the process is in the tail, and with scale $\sigma_{ij}$.
The latent state $z_{ij}$ is a categorical random variate
with unit simplex probabilities
determined by row $z_{ij - 1}$ of the state transition matrix, shown
by the superscript in $\Theta_{ij}^{z_{ij - 1}}$. 
As discussed below, the state transition matrix
may be time-homogeneous or -inhomogeneous, depending on the context.
In the state-transition matrix, 
$\pi$ denotes $p(z_{ij} = 1 \mid z_{ij-1} = 1)$, the 
probability of staying in the body
process at period $j$, and $\nu = p(z_{ij} = 2 \mid z_{ij-1} = 2)$,
the probability of staying in the tail at period $j$.
Their complements, $1 - \pi$ and $1 - \nu$, represent the
probabilities of transitioning from body to tail, and tail
to body, respectively.

The emission distributions, $\mu_{1_{ij}}$ and $\mu_{2_{ij}}$,
used here are two canonical body and tail development models:
the chain ladder model for body loss development factors
\citep{mack1993,englandverrall2002},
and an exponential decay curve following the generalised Bondy model
for the tail process \citep{tailfactors2013}.
The full model specification is completed by choosing these
functional forms, along with the functional form for the variance
and the remaining prior distributions.

\begin{align}
\begin{split}
    \mu_{1_{ij}} &= \log(\alpha_{j - 1} y_{ij-1}) \quad \forall j > 1\\
    \mu_{2_{ij}} &= \log(\omega^{\beta^{j}} y_{ij - 1}) \quad \forall j > 1\\
	\sigma_{ij}^2 &= \exp(\gamma_{1} + \gamma_{2} j + \ln(y_{ij-1}))\\
    \log \bm{\alpha}_{1:M - 1} &\sim \mathrm{Normal}(0, \scriptstyle{\frac{1}{1:M - 1}})\\
    \log \omega &\sim \mathrm{Normal}^{+}(0, 1)\\
    \log \frac{\beta}{1 - \beta} &\sim \mathrm{Normal}(0, 1)\\
    \bm{\gamma}_{1:2} &\sim \mathrm{Normal}(0, 1)\\
\end{split}
\end{align}
Due to the multiplicative autoregressive nature of typical
loss development models, the first data point is not modelled,
and the hidden Markov process is assumed to start in the
body state.
The $M - 1$ body link ratios are given by the vector $\bm{\alpha}$,
and the tail link ratios are given by $\omega^{\beta^{j}}$, for any $j$,
allowing extrapolation to arbitrary future development periods.
The parameter $\omega$ is constained to be strictly greater than $1.0$,
such that growth is monotonically increasing,
and $\beta$ is constrained to lie in the interval $(0, 1)$, to avoid
tail factors growing without bound to $\infty$.
The variance is dependent on development period,
and is proportional
to the losses at the previous time point.

The prior distribution on $\bm{\alpha}$ is regularised towards
a link ratio of 1 in inverse proportion to the development period.
This assumption is imposed because the body link ratios may suffer
from non-identifiabiility at later development periods, when the
latent state is more likely to be in the tail,
$z_{ij} = 2$.
Even when the tail process
is most likely,
some samples could be generated from the body process because
$z_{ij}$ is a random variable, and large values of $\alpha_{ij-1}$
might have unwanted influence on the predictions.
This is akin to the use of pseudo-priors in Bayesian mixture
models \citep{carlinchib1995}, which ensure parameters in the mixture
not being sampled do not become implausible.

\subsection{Model variants}
Three variations of the hidden Markov model are used (Table \ref{table:variants}). 
The base model (referred to as HMM) has homogeneous transition matrix $\Theta_{ij}$,
and sets $\nu$ to zero. This implies that once the tail
process is active, the model cannot switch back to the
body process. This is the primary assumption underlying
tail modelling generally: at some development point, the 
losses smoothly develop to ultimate. Secondly, the HMM-$\nu$
model estimates $\nu$, allowing for the tail to switch
back to the body state. Some triangles may illustrate
unexpected late-development volatility, at which point
the more flexible body process is a better explanation
of the data. Finally, the HMM-lag variant allows $\pi$
to vary across development periods, $\bm{\pi} = (\pi_{1}, 
..., \pi_{j - 1}, ..., \pi_{M - 1})$, on which 
an ordered assumption is imposed such that the probability
of transitioning to the tail increases with development period.
Many extensions of these variants are possible, including
the addition of covariates on the estimation of
$\Theta_{ij}$ or other parametric or non-parametric
forms. This paper restricts focus on these three
variants as they present the simplest use cases
to compare to the two-step process. Although
the HMM-lag and HMM-$\nu$ variants could be combined,
early tests of the hidden Markov development model
by the author indicated that this model risked unidentifiability
without further data or covariates.

\begin{table}
    \centering
    \begin{tabular}{p{2cm}|p{3.5cm}|p{7cm}}
        Name & $\Theta_{ij}$ & Description \\
        \hline\\
        HMM & $\begin{pmatrix} \pi & 1 - \pi \\ 0 & 1 \end{pmatrix}$ & 
        Body-to-tail and body-to-body transition probabilities estimated, constant
        across experience periods.\\
        HMM-$\nu$ & $\begin{pmatrix} \pi & 1 - \pi \\ \nu & 1 - \nu \end{pmatrix}$ &
        All transition probabilities estimated, constant across experience periods.\\
        HMM-lag & $\begin{pmatrix} \pi_{ij-1} & 1 - \pi_{ij-1} \\ 0 & 1 \end{pmatrix}$ &
        Body-to-tail and body-to-body transition probabilities estimated,
        varied across experience periods.\\
    \end{tabular}
    \caption{
        The three hidden Markov model transition matrix variants
        used in the examples.
    }
    \label{table:variants}
\end{table}

\subsection{Estimation}
The models were fit Stan \citep{stan2017} using Bayesian inference
via Hamiltonian Monte Carlo, via the \texttt{cmdstanpy} \citep{cmdstanpy2024} 
Python package and \texttt{cmdstan} \citep{cmdstan2024}
command line interface to Stan, version 2.36.0.
Stan requires specifying a statement
proportional to the joint log density of the data and parameters.
For the above model, the log joint density for a single data point,
grouping all transition matrix parameters as $\bm{\psi} = \{\pi, \nu\}$
and emission distribution parameters as $\bm{\phi} = \{\alpha,
\omega, \beta, \gamma\}$, can be factored as:

\begin{equation}
p(y_{ij}, z_{ij}, z_{ij-1}, \bm{\phi}, \bm{\psi}) = 
    p(y_{ij} \mid  z_{ij}, \bm{\phi})
    p(z_{ij} \mid z_{ij-1}, \bm{\psi})
    p(z_{ij-1} \mid \bm{\psi})
    p(\bm{\psi})
    p(\bm{\phi})
\end{equation}
Stan's use of Hamiltonian Monte Carlo means it cannot
directly estimate the discrete parameters 
$z_{ij}$ or $z_{ij-1}$, meaning they
must be marginalised out of the model statement.
Indeed, marginalisation is often more computationally
efficient and stable than estimating discrete parameters
directly.
However, naive marginalisation schemes are inefficient
because the number of possible latent state paths increases
exponentially with the number of time points,
providing all states can be transitioned between
at each time point. Therefore, the hidden Markov
models were estimated using the forward algorithm
\citep{rabiner1989},
which iteratively stores the joint probability
of the data up to the $j$th development period
in the $i$th accident period and 
being in each discrete parameter possibilities
for easy marginalisation, known as the forward
probabilities. 
The log joint density for a single accident year under
this scheme is:

\begin{equation}
    p(y_{1:M}, \bm{\phi}, \bm{\psi}) = p(y_{iM}, y_{i,1:M-1} \mid \bm{\phi}, \bm{\psi})
    p(\bm{\phi}) p(\bm{\psi})
\end{equation}
where $y_{iM}$ are the losses at the
last development period for accident period $i$
and $y_{i,1:M-1}$ are all losses for accident period $i$ up to 
development period $M - 1$. The joint
density of these two terms marginalises
across latent states using the expression:

\begin{align}
    \begin{split}
        p(y_{iM}, y_{i,1:M-1} \mid \bm{\phi}, \bm{\psi}) =
        &\sum_{k=1}^{K} p(y_{ij} \mid z_{ij} = k, \bm{\phi})\\
        &\sum_{h=1}^{K} p(z_{ij} = k \mid z_{ij-1} = h, \bm{\psi})
            p(z_{ij - 1} = h, y_{i, 1:j})
    \end{split}
\end{align}
where $p(z_{ij-1} = h, y_{i,1:j})$ is the stored
forward probability for the $h$th latent state possibility
up to development lag $j - 1$.
After model fitting, the hidden states on the
training data are recovered using the
Viterbi algorithm \citep{rabiner1989},
which provides the most likely joint
sequence of latent states that generated
the data. For future data, new samples of
$z_{ij}$ are simulated from a categorical
distribution with estimated parameters of
the transition matrix $\Theta_{ij}$ for
that data point.

\section{Competing statistical models}
The hidden Markov development model is compared to
$i$) a slightly more parsimonious latent change-point
model and $ii$) a two-step modelling approach traditional
in actuarial science.
Denote $\tau \in \{2, ..., M\}$ the body to tail
switch-over development period, and
$\bm{\rho} = (\rho_{1}, \rho_{2}) \in \{2, ..., M\}$,
where $\rho_{1} < \rho_{2}$,
a vector of tail start and end training window
development points, respectively.
While these constants could in theory vary over
experience periods, there is typically insufficient
data to do so.

In the traditional two-step modelling process,
the chain ladder
method is first fit to training data up
to development period $\tau - 1$ and
predictions of the lower diagonal loss
triangle are made up to $\tau$, only.
Secondly, the chosen tail model is fit to data
lying within the development period
interval $[\rho_{1}, \rho_{2}]$,
and predictions from $\tau$ made to 
some arbitrary development period, $j^{*}$.
The challenge and art of this two-step process
is in finding a value for $\tau$ that
identifies the development period where
losses are plateauing in the tail,
and finding values for $\bm{\rho}$ that
identify a suitable decaying curve of link ratios.
While $\tau = \rho_{1}$ in some cases,
more generally $\rho_{1} \leq \tau$.
The two-step approach differs from
equation (\ref{eq:hmm}) in only two
ways:

\begin{align}
\begin{split}
	y_{ij} &\sim 
	\begin{cases}
		\mathrm{Lognormal}(\mu_{1}, \sigma_{ij}) \quad j < \tau\\
		\mathrm{Lognormal}(\mu_{2}, \sigma_{ij}) \quad \rho_{1} \leq j \leq \rho_{2}
	\end{cases}\\
	\log \bm{\alpha}_{1:\tau - 2} &\sim \mathrm{Normal}(0, 1)\\
\end{split}
\end{align}
where now the decision between the two
models is decided by $\tau$ and $\bm{\rho}$.
The further exception in the two-step approach
is the use of a standard normal prior
on the log-scale link ratios, rather than the
regularising prior used in the hidden Markov models,
because non-identifiability is, in general,
no longer a concern.

The latent change-point model offers a compromise
between the two-step modelling process
and the hidden Markov model.
Like the hidden Markov model, the chain ladder
and generalized Bondy models are fit jointly, 
and like the two-step process the models
switch over at some development period
$\tau$. However,
$\tau$ is considered to be a
discrete parameter estimated
in the model directly.
The $\bm{\rho}$ parameters are not necessary
because the generalized Bondy process is
only fit from $\tau$ onwards.
In practice, due to its discrete nature,
$\tau$ is marginalised out of the model,
and the prior on $\tau$, $p(\tau)$,
is a discrete uniform distribution over the possible
development period switch-over points in $\{2,...,M\}$.

\begin{align}
\begin{split}
	y_{ij} &\sim 
	\begin{cases}
		\mathrm{Lognormal}(\mu_{1}, \sigma_{ij}) \quad j < \tau\\
		\mathrm{Lognormal}(\mu_{2}, \sigma_{ij}) \quad j \geq \tau\\
	\end{cases}\\
    \log \bm{\alpha}_{1:M - 1} &\sim \mathrm{Normal}(0, \scriptstyle{\frac{1}{1:M-1}})\\
    \tau &\sim \mathrm{DiscreteUniform}(2, M)\\
\end{split}
\end{align}
Like the hidden Markov model, we place a regularising
prior on the log-scale link ratios of the chain-ladder
component to handle potential non-identifiability
at higher development periods where the process
is more likely to be in the tail.

\section{Datasets and model performance}
We compared the hidden Markov development model to
the latent change-point and two-step models 
using two sets of data.

\begin{figure}
    \centering
    \includegraphics[scale=0.4]{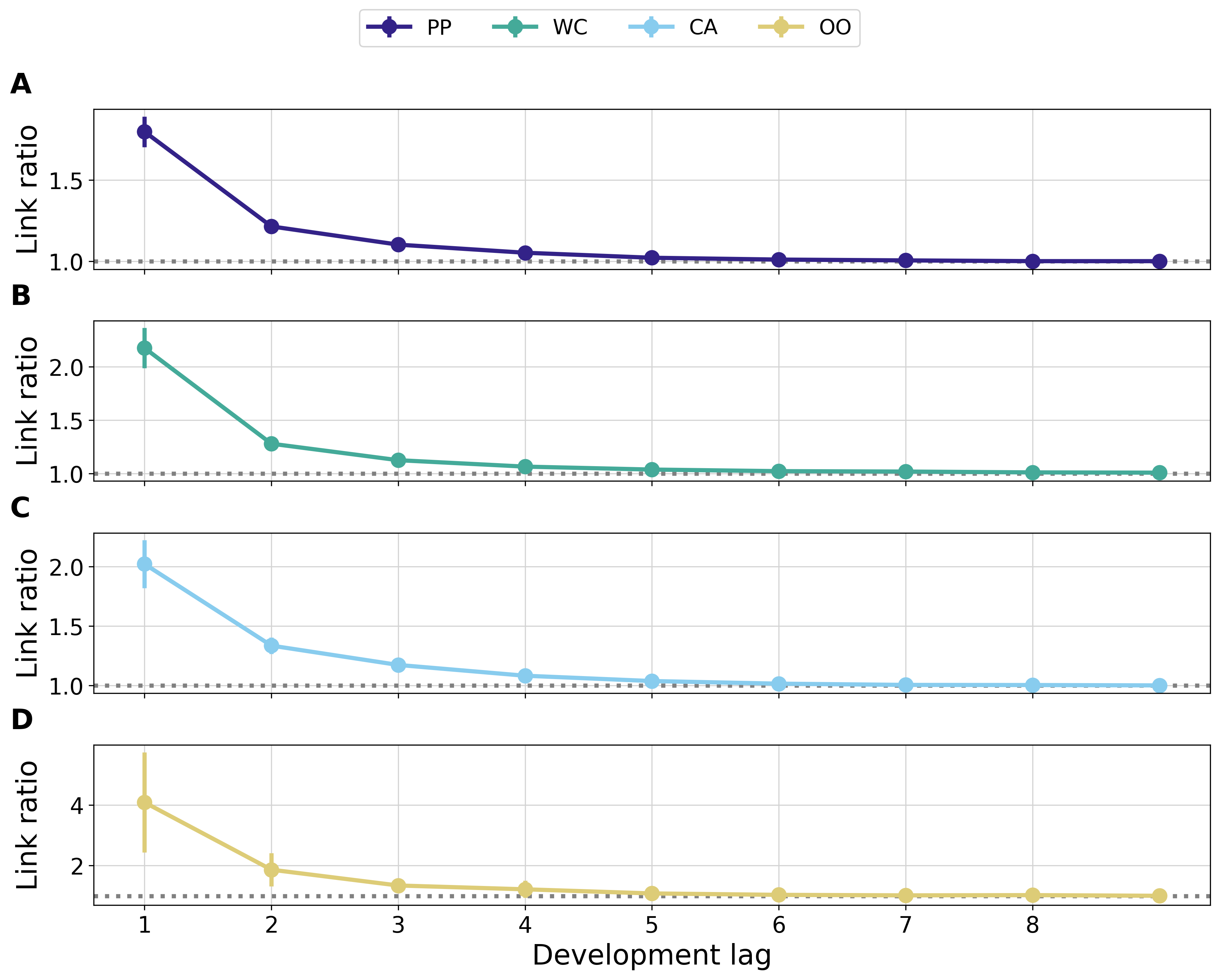}
    \caption{
        The empirical link ratios by line of business 
		(panels A-D) in
        the \cite{meyers2015} dataset.
        Points indicate the mean across triangles,
        and vertical line segments show 2 standard
        deviations.
    }
	\label{fig:industry-atas}
\end{figure}

Firstly, the models were fit to 200 industry
paid loss triangles from \cite{meyers2015}, 50 triangles
for the four lines of business: private passenger auto (PP),
workers compensation (WC), commerical auto (CA), and
other occurrence (i.e. general) liability (OO).
We removed three triangles with zero loss values first,
one from the CA dataset and two from the OO dataset.
Each triangle covered 10 years of historical
accident and development periods, amenable to partitioning
into an upper diagonal of training data,
$\mathcal{Y}$, and lower diagonal of test data,
$\tilde{\mathcal{Y}}$.
As these are yearly triangles, and have been
chosen for previous model validation exercises \citep{meyers2015},
it is reasonable to assume that the losses at development
period 10 are close to ultimate.
For the two-step approach, we inspected the
mean and standard deviations of the empirical
link ratios across triangles (shown in Figure
\ref{fig:industry-atas}), and selected
$\tau = 6$ and $\bm{\rho} = (4, 10)$. These
were chosen given
that the link ratios showed smooth patterns
of decay from approximately development period
4 onwards, and the
triangles were close, on average, to their values
at period 10 by development period $\tau = 6$.

When fitting the models to industry triangles,
a small number of models
produced
very large posterior predictions on out-of-sample
data, which numerically overflowed. The multiplicative
autoregressive nature of typical loss development models
mean that large predictions at one
time point can quickly compound to unrealistic and
computationally-unstable values. For this reason,
we capped the predictions at 100 times the
maximum value across the training and test data
for each triangle.

Secondly, we used 5
triangles presented in the
relatively recent literature on key
papers on loss development
modelling,
including tail development, that provide more
historical data than the industry triangles:
the long-tailed liability and short-tailed property 
quarterly triangles from medium-size insurers
in \cite{balona2022}, the annual liability triangle in \cite{merz2015},
the Swiss annual liability triangle from \cite{gisler2009},
and the annual liability triangle in \cite{verrall2012}.
These triangles had between 17 and 22 periods of
data. For each triangle, we used the latest diagonal as the
test dataset to evaluate predictive performance.
For the two-step modelling approach, we chose the
$\tau$ and $\bm{\rho}$ constants
by visual inspection of the empirical link ratios,
selecting $(4, 4, 4, 4, 11)$ for $\tau$ values
for each triangle,
and $[(4, 16), (3, 20), (3, 21), (3, 16), (10, 21)]$
for $\bm{\rho}$ values, respectively.
The link ratios are displayed in Figure \ref{fig:literature}.

\subsection{Model performance}
Model performance can be split into two facets:
accuracy and calibration.
We summarised model accuracy using two metrics
applied to future out-of-sample data: the expected
log predictive density (ELPD) and the root mean
square error (RMSE). 
The ELPD \citep{vehtari2017} is based on the logarithmic scoring
rule, and is defined for a single triangle
as the joint log density of the out-of-sample
data and posterior distribution of the parameters. 
In this way, ELPD measures the log probability density
of the out-of-sample data from the estimated model,
and more heavily
penalises models further from
the true data generating process.
Across each accident period $i$ and development
period $j$ in $\mathcal{\tilde{Y}}$, we take the
sum of log likelihood values marginalized across the posterior
samples:

\begin{align}
	\label{eq:elpd}
	\begin{split}
	\mathrm{ELPD} &= \sum_{i=1}^{N} \sum_{j=N - i + 1}^{M} 
					\log p(\tilde{y}_{ij} \mid \mathcal{Y})\\
				 &=	\sum_{i=1}^{N} \sum_{j=N - i + 1}^{M} 
					\log \int p(\tilde{y}_{ij} \mid \theta)
					p(\theta \mid \mathcal{Y})
					d \theta \\
				 &\approx \sum_{i=1}^{N} \sum_{j=N - i + 1}^{M} 
					\frac{1}{S} \sum_{s=1}^{S} 
					\log p(\tilde{y}_{ij}^{(s)} \mid \mathcal{Y})
	\end{split}
\end{align}
where $p(\tilde{y}_{ij} \mid \mathcal{Y})$
is the posterior predictive distribution for
the $i$th accident period at lag $j$,
$\theta$ is used to generically refer to all
model parameters, i.e. $\theta = \{\bm{\phi}, \bm{\psi}, \bm{z}\}$,
and the super-script in $\tilde{y}_{ij}^{(s)}$ denotes the
$s$th sample from the posterior distribution with $S$
total samples.
The second sum over $N - i + 1$ development periods in the
$i$th accident period assumes a typical loss triangle
with a full lower diagonal of test data.
To compare models, we
calculated the difference in ELPD for each triangle, $t$,
and its standard error,
where the standard error of the difference is the
square root of the product of $i$) the sample variance of log predictive
density differences between models, and $ii$) 
the number of data points \citep{vehtari2017,sivula2020}.
For the industry data, we combined ELPD values for each
triangle by taking the mean of the ELPD differences and 
mean of the standard errors \citep{sivula2020}. 
Approximate 95\% confidence intervals were then derived
by using a range of 2 standard errors around the estimate.
Although the test data is known with certainty,
it is still just a portion of accident periods
and development lags
for each triangle, and therefore uncertainty
in ELPD was still calculated.

The RMSE was defined per out-of-sample data point in
$\tilde{\mathcal{Y}}$ as:

\begin{equation}
	\label{eq:rmse}
	\mathrm{RMSE_{ij}} = \sqrt{\frac{1}{S} \sum_{s=1}^{S} (\hat{y}_{ij}^{(s)} - \tilde{y}_{ij})^2}
\end{equation}
where $\hat{y}_{ij}^{(s)}$ is the $s$th sample from the 
posterior predictive distribution.
In contrast to ELPD, RMSE penalises models that produce
predictions further from the test data points using a
quadratic scoring rule,
and may demonstrate different results depending
on the context.
As with ELPD, the average differences in RMSE
between models per triangle were used to compare models, and 
their standard errors were derived as the square root
of the product of $i$) the sample variance of the differences
in RMSE and $ii$) the number of data points.

Model calibration was inspected using histograms
of the percentiles of the true data on the 
posterior predictive distributions, i.e.
the empirical cumulative distribution functions, 
a quantity sometimes referred to as the
randomised quantile residual \citep{dunn1996}.
Well-calibrated models' percentiles should be
approximately uniformly distributed.

\section{Simulation-based calibration and simulated examples}
To validate the hidden Markov model implmentation, 
we use both simulated examples
to build intuition, alongside
simulation-based calibration \citep{talts2018,modrak2023}.
Simulation-based calibration leverages the self-consistency
of the Bayesian joint distribution of parameters and
data: fitting a Bayesian model to datasets generated
from its prior predictive distribution and averaging 
over datasets should return the prior distribution.
Specific simulated examples, in addition, helps
build intuition about model dynamics.

\subsection{Simulation-based calibration}
Focusing on the base HMM model variant, 1000 
full-triangles with $N = M = 10$ were generated
from the prior predictive distribution, and the
HMM model fit to each upper diagonal $\mathcal{Y}$.
The prior distributions were the same as in
equation (\ref{eq:hmm}), except for the priors
on $(\gamma_{1}, \gamma_{2})$, which were 
given more informative normally-distributed priors
with locations and scales of $(-3, -0.25)$ and
$(-1, 0.1)$, respectively. Due to the
multiplicative autoregressive forms in the
location and scales of the likelihood in
equation (\ref{eq:hmm}), particularly
large values for $\sigma$ can cause overflow
in the sampled data.

Each model was summarised by
calculating the rank statistics of quantities
of interest.
The rank statistic is the number of times a
simulated value is greater than the posterior values,
and should be approximately uniformly distributed
if the model has been implemented correctly and is
unbiased \citep{talts2018}. To reduce the autocorrelation
in the posterior distributions, the posteriors were
thinned to every 10th posterior draw.
Rank statistics
were calculated for each parameter in the HMM model,
as well as the joint log likelihood and
an the ultimate loss prediction at data point
$(i=1, j=10)$, since \cite{modrak2023} recommend
using test quantities that average over the entire
data space in evaluating SBC.

\begin{figure}
    \centering
    \includegraphics[scale=0.45]{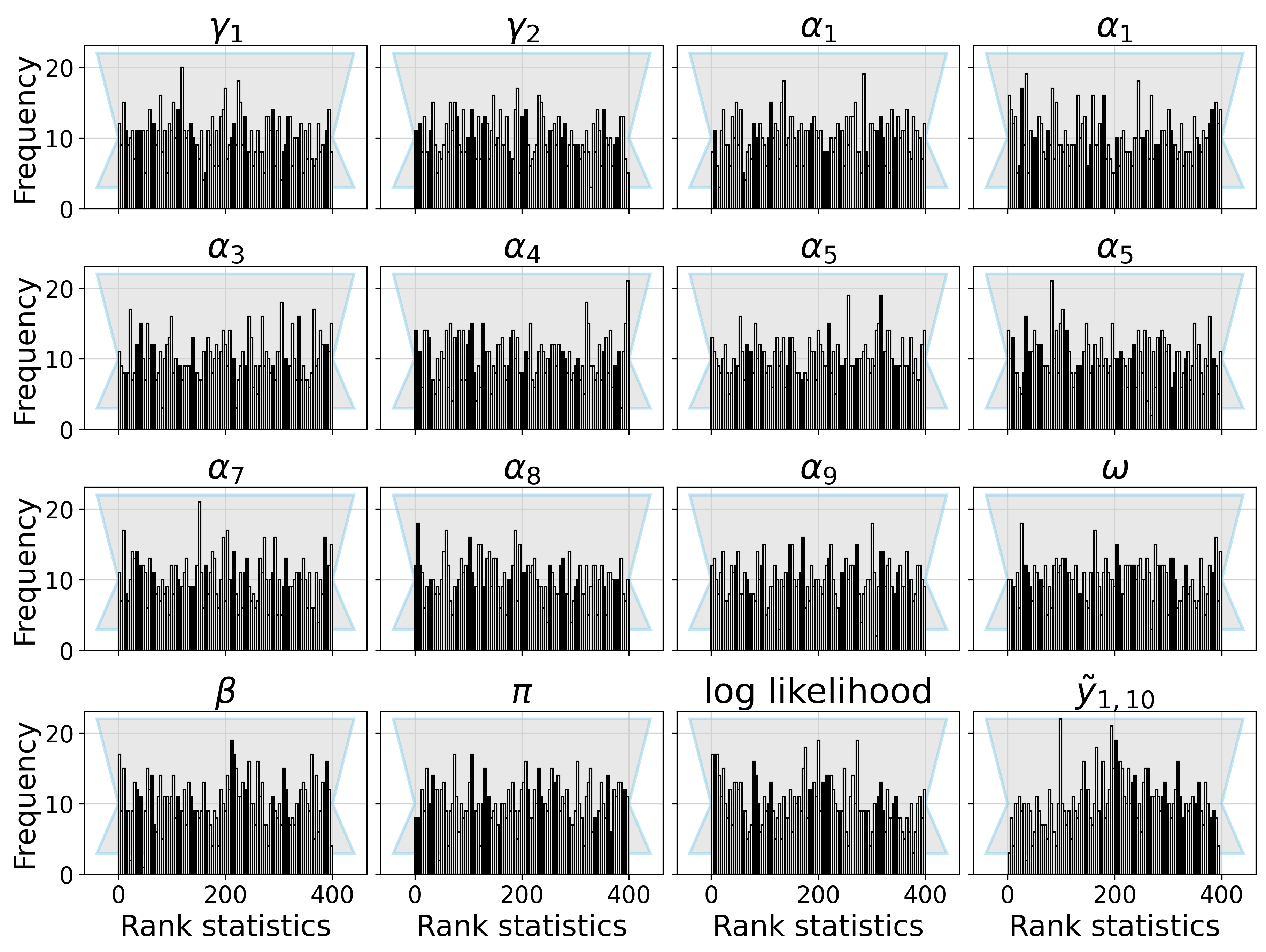}
    \caption{
		Histograms of simulation-based calibration
		rank statistics for the HMM model, 
        with the 99\% percentile
        interval from a discrete uniform distribution
        shown in the grey shaded band.
        Each histogram shows a key model parameter,
        and the final two panels show the ranks
        for the joint log
        likelihood and the first ultimate loss
        distribution.
        For each model, we sampled
        4000 draws from the posterior distribution,
        and thinned the samples by 10 to remove any
        autocorrelation, meaning a maxmimum rank statistic
        of 400. Of the 1000 models,
        10 were removed due to poor convergence.
    }
	\label{fig:sbc}
\end{figure}

No problems with model calibration were found using simulation-based
calibration (Figure \ref{fig:sbc}),
with all histograms matching the assumptions of
uniformity. However, 10 models were removed for poor
convergence, which typically occurred when the simulated
link ratios from the body process were
higher than values expected from real loss triangles. 
Given this occurred rarely, the priors were left
unchanged, although suitable prior distributions
for Bayesian chain ladder models is an area
with a dearth of literature.
For the 990 models, 
the average classification accuracy of the recovered latent
state values $\bm{z}$, across
both training and test data, was 97\% with a 
95\% highest density interval (i.e. the 95\% most
likely values) of [91, 100].

\subsection{Simulated example}
\begin{figure}
    \vspace{-1.25cm}
    \centering
    \includegraphics[scale=0.45]{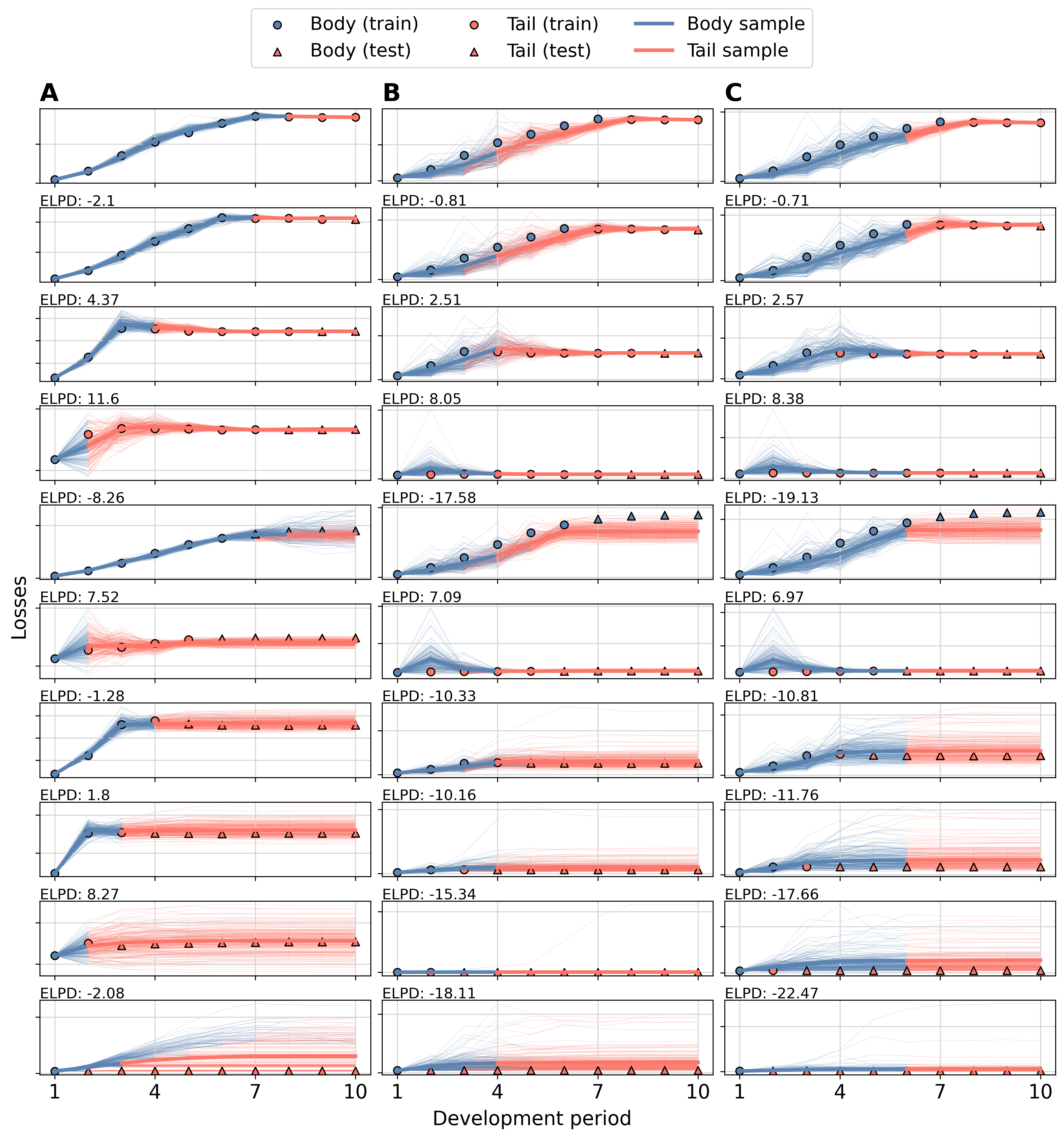}
    \caption{
        Comparison of the HMM (panel A), 
        latent change-point (panel B)
        and two-step (panel C) models on ten accident
        periods (rows) of data simulated from
        the HMM model.
        The true losses
        are shown as points, with colours
        identifying true
        body (blue) or tail (orange) status.
        Circles denote training data and
        triangles test data. Thin 
        lines are samples from the posterior
        predictive distributions, coloured
        by latent state, and the thicker
        line shows the posterior mean.
        The ELPDs on the test data
        points in each experience period
        are shown above each plot.
        The two-step model was fit
        assuming $\tau = 6$ and
        $\bm{\rho} = (6, 10)$.
    }
	\label{fig:numerical}
\end{figure}

To build intuition for the two models,
Figure \ref{fig:numerical} shows a numerical
simulation of data from a HMM model,
and the results for the HMM model
(panel A), the latent change-point
model (panel B), and the
two-step model (panel C).
As shown by the ELPD values above each
plot with available test data (experience
periods two to ten), 
the HMM variant outperforms
the latent change-point and two-step approach
in all
experience periods except the second.
For the two-step process, the chosen value $\tau = 6$
matches the HMM data-generating process
closely for that experience period, 
resulting in higher accuracy. 
The latent change-point model detected the most
likely switch-over point at $\tau = 4$ with probability
46.5\%, but the relatively smooth dynamics of the
second experience period and 9 training data points
results in accurate estimation and predictions. 
In the remaining experience
periods, however, the two-step and latent change-point approaches
generalise poorly.
For instance, the third experience period (third row)
shows that the consequences of the $\tau$ values results
in over-estimation of losses in the latent change-point
and two-step approaches.
Similarly, the fifth experience period (fifth row)
shows a case where the tail model is never
entered by the HMM process, and the
latent change-point and two-step processes
consequently under-estimate true losses.
This example illustrates that
if $\tau$ is chosen correctly, there is sufficient
data available for model training,
and the data is typically well-behaved,
the two-step and latent change-point approaches 
can provide more exact predictions.
However, if experience periods differ in their
body-to-tail switch-over dynamics,
which is highly expected, then overall
performance suffers due to
growing generalisation error.

\section{Performance on industry data}
\label{sec:results}
\begin{figure}
    \centering
    \includegraphics[scale=0.45]{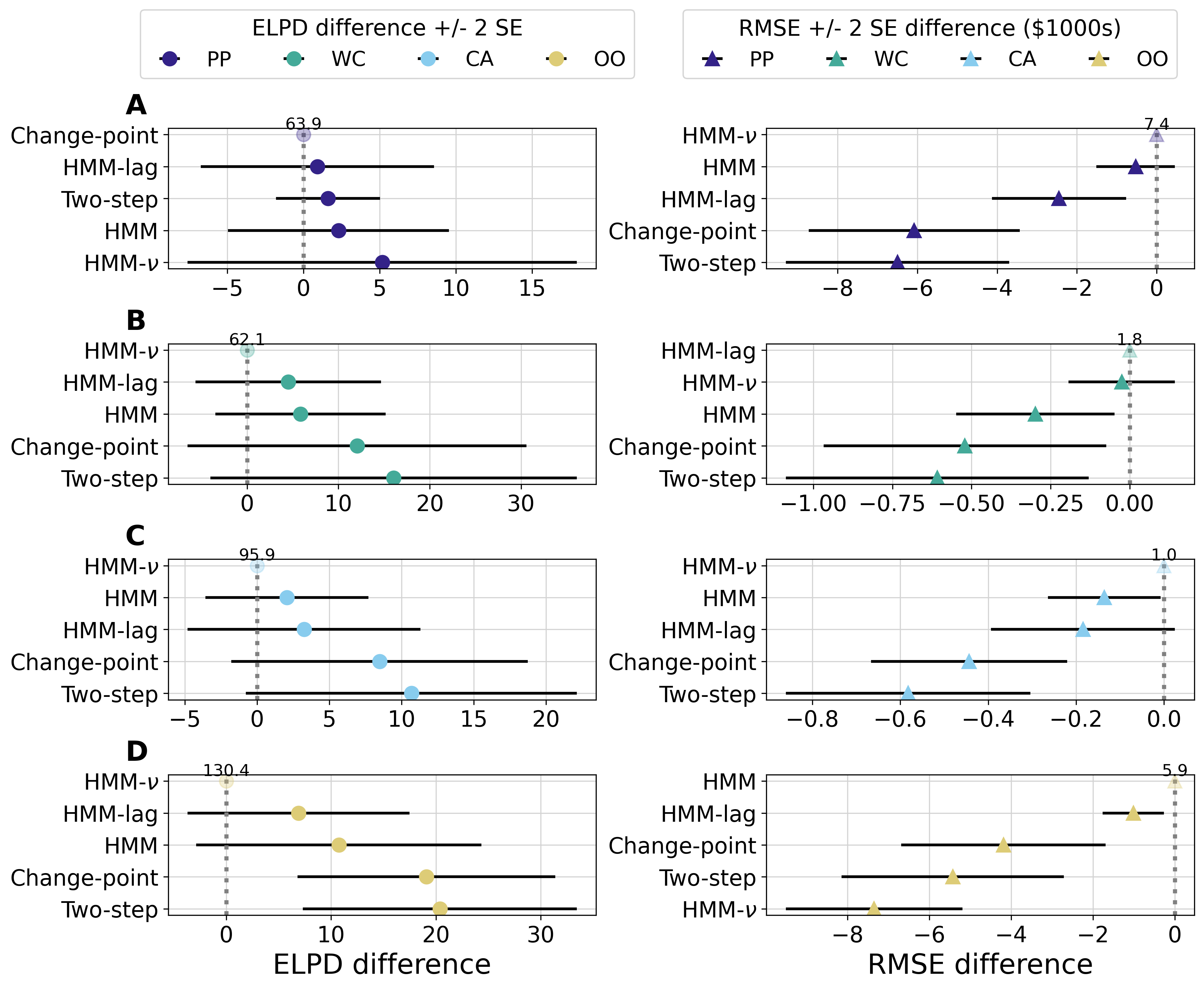}
    \caption{
        The ELPD (left column) and RMSE (in thousands of dollars; right column)
		differences (+/- 2 standard errors; SE) in order of performance
        for each model and line of business for the industry
        triangles. Rows A through D enumerate results
        for line of business separately.
        The best-performing model is shown at the top of each
        panel, with the absolute ELPD or RMSE value displayed above.
        Positive ELPD differences with an uncertainty interval that does not
        cross zero indicates a credible difference at the 95\% level
        in favour of the top model.
        Negative RMSE differences with an uncertainty interval
        that does not cross zero indicates a credible difference
        in favour of the top model.
    }
	\label{fig:backtest-scores}
\end{figure}

The ELPD and RMSE differences between models, relative to
the best model,
are shown in Figure \ref{fig:backtest-scores}.
When calculating ELPD, a small number of pointwise log densities 
showed very small, negative values, indicating poor predictions
on the out-of-sample data leading to numerical instability. 
We decided to remove any log predictive
densities for all models that had values $< -100$, which for a single
out-of-sample data point was particularly low, given that most
ELPD values for a single data point lie within [-5, 5]. This retained 99.7\%
of values for the PP results, 99.1\% of values for WC, 99.6\%
for CA, and 98.2\% for OO. The full log densities are given in the
supplementary materials, as well as results for different levels
of filtering.

For ELPD, at least one of the
hidden Markov model variants out-performed
the latent change-point and 
two-step models, on average, 
apart from ELPD values for the PP
lines of business where the latent change-point
performed marginally better than the nearest
HMM model. The standard errors around the ELPD
values, however, indicated that all the models
often performed within 2 standard errors from each other,
apart from the OO line of business where the HMM models
performed noticeably better.
For all lines of business, one of the HMM 
models clearly out-performed the latent change-point
and two-step models in RMSE.
Overall, the HMM-$\nu$ model attained
75\% of the best average ELPD scores, and 50\%
of the best average RMSE scores, meaning that
allowing for tail processes to revert
to the body is an important feature in making
future predictions.
Evaluating the predictions at the 10th
development period
out-of-sample values only, to mirror
the estimation of ultimate loss,
mirrored the results in Figure
\ref{fig:backtest-scores} apart from some
minor changes in ranks of the HMM
models (full results are supplied
in the code repository).
The performance results also indicated
that the latent change-point model
was close to, but still out-performed,
on average, the two-step in both ELPD and RMSE.

\begin{figure}
    \centering
    \includegraphics[scale=0.45]{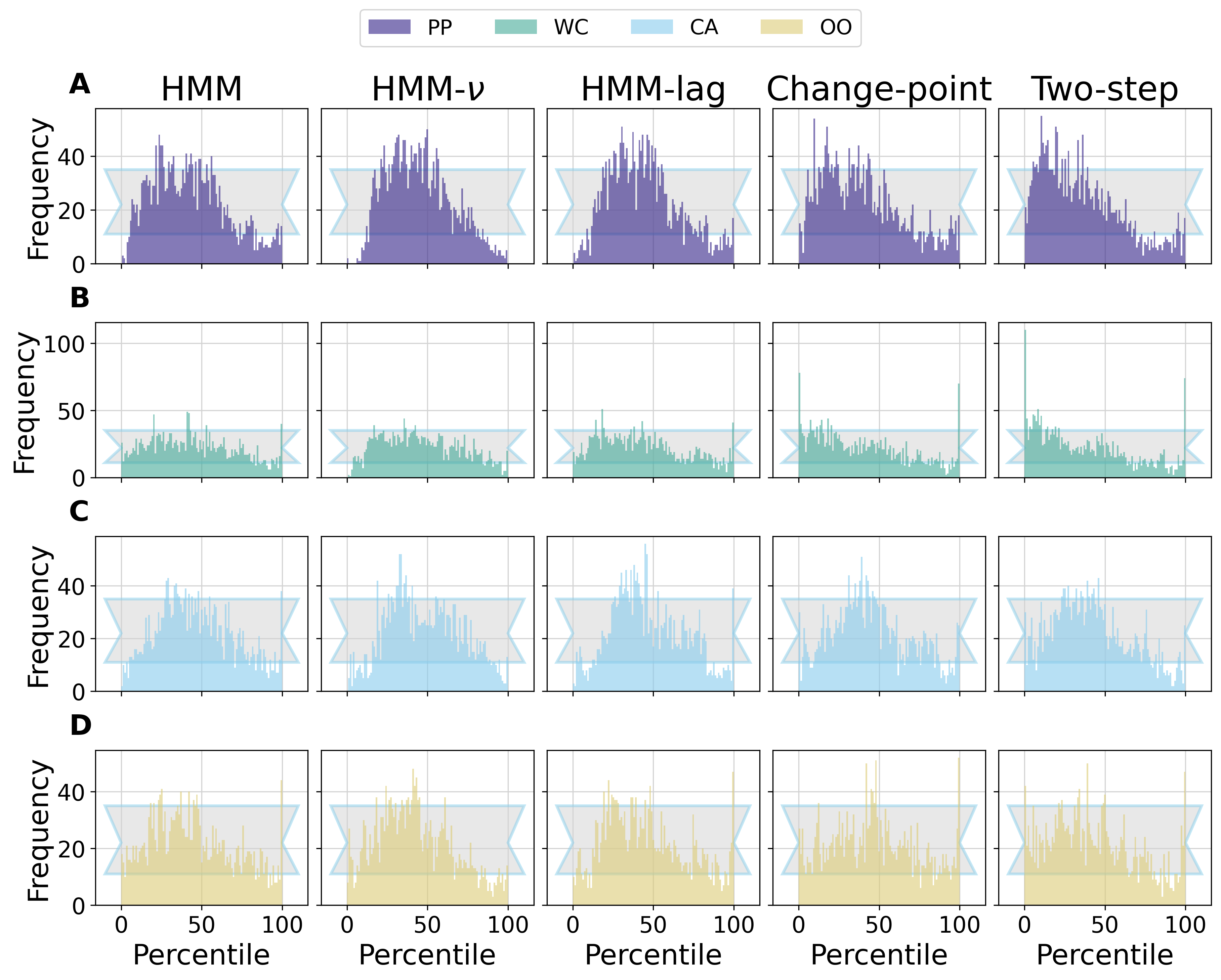}
    \caption{
        Percentiles of the true left-out values on the posterior distributions
        for each model and line of business (panels A through D) in the industry triangles.
        Grey shaded regions provide the 99\% intervals of a discrete uniform
        distribution, for reference.
        Right-skewed histograms indicate under-estimation,
        left-skewed histograms indicate over-estimation,
        and inverted-U histograms indicate predictions that
        are uncertain.
    }
    \label{fig:percentiles}
\end{figure}

Model calibration histograms (Figure \ref{fig:percentiles})
indicated that both the hidden Markov models
and two-step approaches typically have predictions that
are too uncertain, indicated by the inverted-U shaped
histograms. In WC, the hidden Markov models had the most
uniform percentiles, whereas the two-step approach
showed signs of both under- and over-estimation.

\begin{figure}
    \centering
    \includegraphics[scale=0.45]{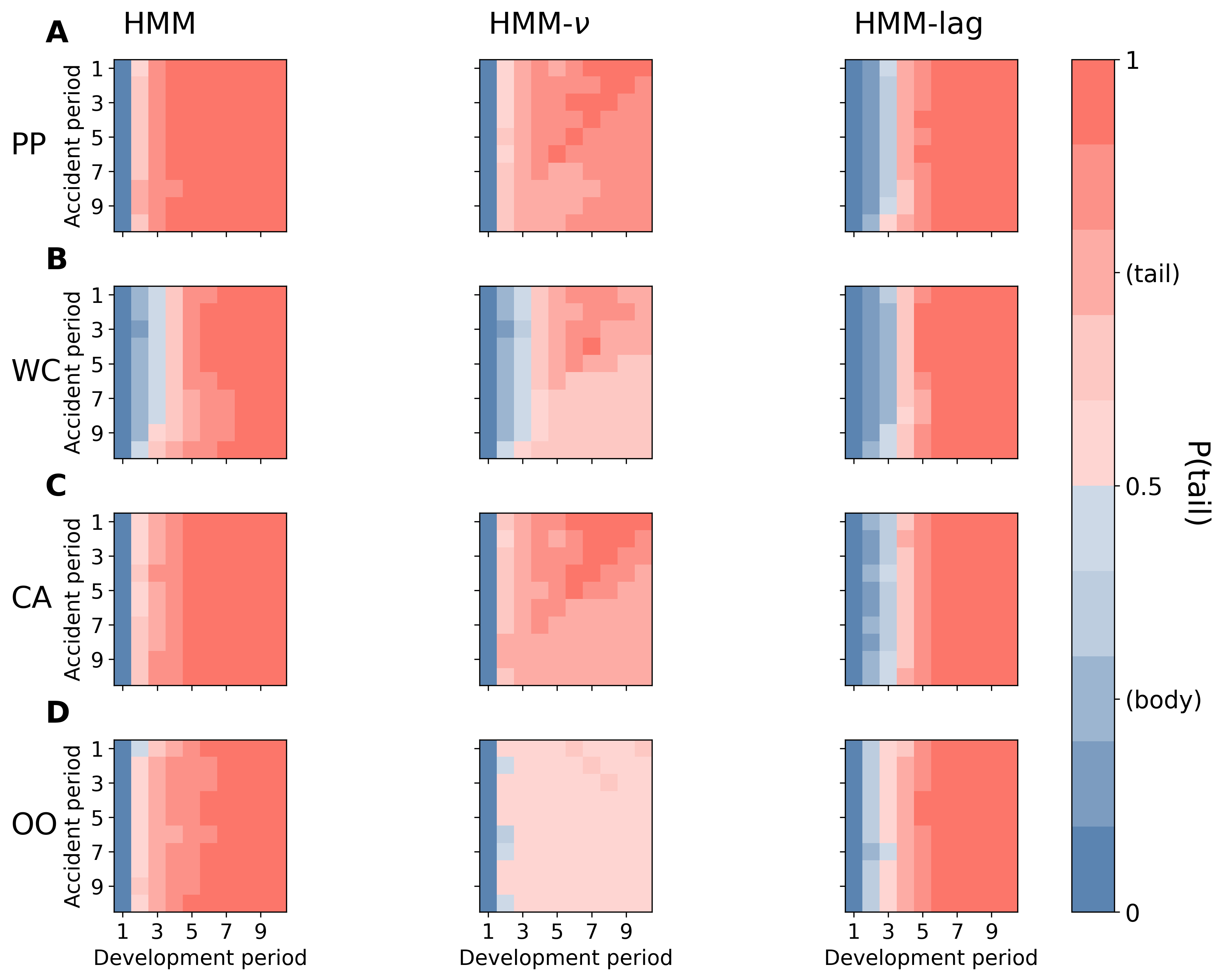}
    \caption{
        The average probability of being in the tail process
        for each hidden Markov model parameterization (columns)
        and line of business (rows A-D) across triangles
        in the industry data.
        Probabilities $\leq 0.5$ are coloured
        in blue whereas probabilities $> 0.5$
        are coloured in orange. More faded squares
        indicating smaller probabilities of being
        in body and tail processes, respectively.
    }
    \label{fig:zstars}
\end{figure}

The hidden Markov model variants had different
implications for body-to-tail switch-over points
depending on the particular line of business
(Figure \ref{fig:zstars}).
The PP and CA lines showed, in general,
the quickest development to the tail
state, at development period 2 for the HMM
and HMM-$\nu$ models, and by development
period 3 for most accident periods for the
HMM-lag model. In contrast, WC stayed
longest in the body state, followed by
OO, and both WC and OO lines demonstrated
relatively equitable probabilities of being
in the body and tail at later development
periods. This is particularly noticeable for
the HMM-$\nu$ model, where the chance
of returning to the body from tail process
was allowed.

\begin{figure}
    \centering
    \hspace{-1em}
    \includegraphics[scale=0.4]{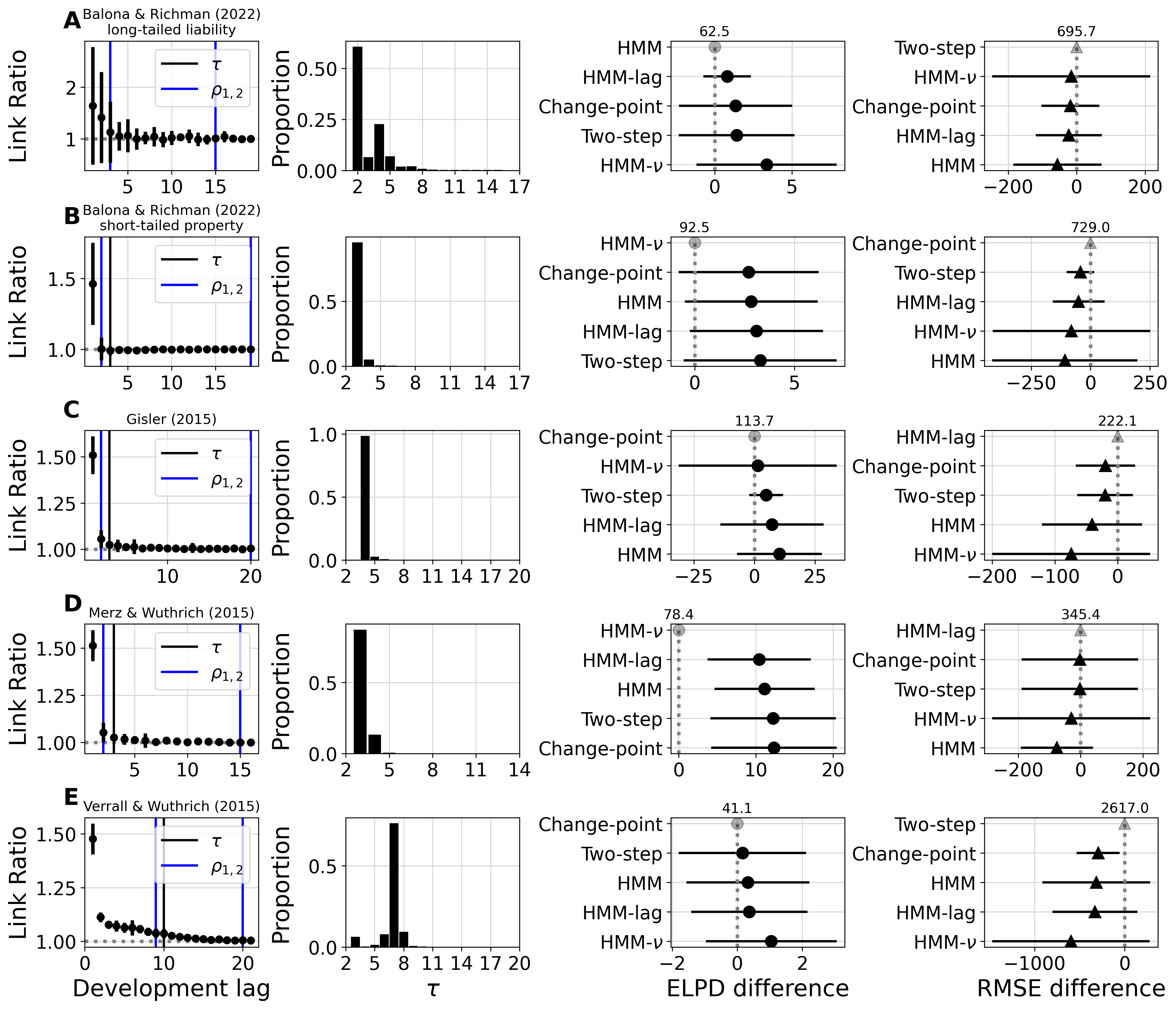}
    \caption{
        Model results on the five literature triangles
        (rows A - D). The first
        column shows the mean (+/- 2 standard deviations)
        of the empirical link ratios in the triangles.
        The black and blue vertical lines indicate $\tau$
        and $\bm{\rho} = \rho_{1:2}$, the tail start
        and generalised Bondy model training windows,
        respectively (see model definitions, above).
        The second column presents the distribution
        of $\tau$ values from the latent change-point,
        for comparison to the two-step manual values
        shown in the first column.
        The third and final columns show the
        ELPD and RMSE differences (+/- 2 SE) from the best
        performing model (top model in each panel), calculated
        from predictions on the latest diagonal of
        data (out-of-sample) in the loss triangles.
        The absolute ELPD and RMSE of the best-performing
        models are shown above the top model, for reference.
    }
	\label{fig:literature}
\end{figure}

For the five literature triangles, the lack
of hold-out data meant that the uncertainties
around the ELPD and RMSE differences
indicated more equal model performance
(Figure \ref{fig:literature}).
The average ELPD
differences indicated that one of the hidden
Markov models or the latent change-point
model performed better than the
two-step approach.
However, the two-step model
performed on-par with the change-point
and hidden Markov models in terms of RMSE,
and clearly performed best in RMSE
for the \cite{verrall2015} triangle
because of the triangle's relative smooth, 
gradual dynamics.
The manual selection of $\tau$
for the two-step process often
closely aligned with the 
posterior distribution of most
plausible $\tau$ values estimated by the latent
change-point model.

\section{Discussion}
\label{sec:discussion}
This paper has proposed
a hidden Markov loss development model
for insurance loss triangles that combines
body and tail development models,
and automates the selection of body-to-tail
switch-over points.
Simulation-based calibration validated
the hidden Markov model implementation
as being unbiased, and across
a range of different datasets,
the hidden Markov model variants
provided similar results to, 
and often out-performed,
the traditional two-step approach
and a latent change-point model.
The hidden Markov models'
automated detection
of body and tail processes
more gracefully captures
loss development dynamics
that may vary over triangle
experience periods, which
reduces analysts' degrees of freedom
that make the traditional
two-step approach reliant
on difficult-to-reproduce
and variable subjective
decisions. The hidden Markov
model further saves analysts'
time in manually selecting
body and tail switch-over points.

The hidden Markov development model 
posits a clear data-generating process for
loss development dynamics.
Although referred to here as `body' and `tail',
these two latent states might equivalently
be thought of as flexible and smooth
periods of loss development, and can interchange
depending on the context, as in the 
HMM-$\nu$ model variant here.
In this way, the hidden Markov model is not a strictly
analogous implementation of the two-step
approach,
as the two-step approach allows analysts
to choose tail model training data windows
that overlap the body-to-tail switch-over point.
Thus, the same data points may be used in
estimation of body and tail processes, rather
than a discrete switch-over point between the two.
Despite this flexibility, the two-step approach
is not a single generative model, and should an
analyst choose values for $\tau$ and $\bm{\rho}$
that are not representative of a particular
experience period, the predictions from
such a model could be extremely biased,
as shown in the example of Figure \ref{fig:numerical}.
While the hidden Markov model may still
make relatively poor predictions for those experience
periods with little data (e.g. see the last
row of panel A in Figure \ref{fig:numerical}),
uncertainty in the true latent state, $\bm{z}$,
is more-accurately accounted for.

Although most of the hidden Markov model variants
performed consistently better on average
than the two-step
approach on the curated industry datasets
of \cite{meyers2015}, the approaches were
more similar on the five literature triangles.
These two sets of data present different
case studies. The industry triangles
have been selected to encompass relatively
large insurers with mostly stable
loss dynamics \citep[see][appendix A]{meyers2015}
over a period of 10 years.
Due to the number of triangles, the two-step
approach's manually-selected variables,
$\tau$ and $\bm{\rho}$,
were chosen based on average empirical link
ratios, which might not have been the best
selection points for some triangles.
By contrast, the literature triangles encompass
more accident periods per triangle but also smaller
books of business \citep[e.g. the medium-sized
triangles from][]{balona2022}, and
more variability in the tail than 
present in the industry triangles.
Previous papers on loss development models
combining body and tail dynamics
have not considered the breadth of triangles
and lines of business used here. For instance,
\cite{englandverrall2001}, \cite{verrall2012},
and \cite{verrall2015} all
used a single triangle to illustrate their approaches,
and 
\cite{zhang2012} used a dataset of
10 workers' compensation triangles and
did not consider other lines of business or
more volatile triangles. Moreover,
the previous papers did not compare
their approaches to the more common
two-step approach applied in actuarial
practice.
The datasets used here are provided alongside
this article in the repository
for ease of access
and comparison of other loss development
modelling approaches.

All models demonstrated relatively poor
calibration on the out-of-sample
data from the \cite{meyers2015}
dataset (Figure \ref{fig:percentiles}). 
Primarily, 
the out-of-sample
predictions were often too
uncertain, producing a
predominance of percentiles falling
closer to 50\% than in the extremes
of the distributions.
Few articles have shown calibration
plots from fully-Bayesian posterior
distributions on out-of-sample
loss development data, so this pattern of calibration requires
further inspection in the literature.
\cite{meyers2015} reports on 
calibration using the same data set,
showing relatively well-calibrated
predictions. However, we note
that \cite{meyers2015} calculated
the percentile of the total ultimate
losses in each triangle on a lognormal
distribution with mean and variance
informed by the total ultimate losses
from their models. Thus, these
are not directly comparable because
the approach in this paper marginalises the percentiles
over the each posterior predictive distribution
in the test data, rather than using the
the mean posterior prediction only.

The hidden Markov models presented here
could be extended in a number of useful
ways. Notably, the transition matrix
probabilities might be parametric
or non-parametric functions of covariates,
such as premium volume in each experience
period or inflation levels in each
calendar period, or include
hierarchical effects for experience
and development periods.
The Bayesian framework, alongside the
hidden Markov models implemented
here and available in the supplementary
material, make these extensions
highly accessible.
Additionally, the hidden Markov
model framework is general enough
to include any body or tail
model, not just the chain ladder
and generalised Bondy forms.
For instance, there are a
number of inverse power curves
to use for tail modelling
\citep{tailfactors2013,evans2015,clark2017},
and extensions and variations on the chain ladder
model have been commonplace \citep{englandverrall2002}.
Although the analyses in this paper focused on
paid losses, the same models could be applied
to estimates of reported losses (i.e.
paid loss plus estimates of reserve),
or joint modelling of both paid and reported
losses \citep[e.g. see][for one approach]{zhang2010}.

\section{Competing interests}
\label{sec:interests}
The author declares no competing interests.

\bibliographystyle{apa}
\bibliography{\src/references.bib}
\end{document}